# Local probing of the field emission stability of vertically aligned multiwalled carbon nanotubes

F. Giubileo[*], A. Di Bartolomeo, A. Scarfato, L. Iemmo, F. Bobba, M. Passacantando[a],

S. Santucci[a] and A. M. Cucolo

*CNR-INFM Laboratorio Regionale SUPERMAT and INFN Gruppo collegato di Salerno and Dipartimento di Fisica "E.R. Caianiello" and Centro Interdipartimentale NanoMateS, Università degli Studi di Salerno, via S. Allende, 84081 Baronissi (SA), Italy*

[a]*Dipartimento di Fisica, Università dell'Aquila and INFN, via Vetoio 67010 Coppito (AQ), Italia*

Abstract:

Metallic cantilever in high vacuum atomic force microscope has been used as anode for field emission experiments from densely packed vertically aligned multi-walled carbon nanotubes. The high spatial resolution provided by the scanning probe technique allowed precise setting of the tip-sample distance in the submicron region. The dimension of the probe (curvature radius below 50nm) allowed to measure current contribution from sample areas smaller than $1\mu m^2$. The study of long-term stability evidenced that on these small areas the field emission current remains stable (within 10% fluctuations) several hours (at least up to 72 hours) at current intensities between $10^{-5}$A and $10^{-8}$A. Improvement of the current stability has been observed after performing long-time Joule heating conditioning to completely remove possible adsorbates on the nanotubes.

[*] Corresponding author: Dr. Filippo Giubileo, e-mail:giubileo@sa.infn.it,
Tel: +39.089.965285, Fax: +39.089.965275

# 1. Introduction

At present carbon nanotubes (CNTs) are considered among the most popular candidates for large-area cold cathode applications because of their simple fabrication and patterning techniques as well as low threshold voltages for field emission (FE) and high enhancement factor. These features make CNTs attractive for a variety of vacuum microelectronic applications: field emission lighting elements [1], field emission flat panel displays [2], gas discharge tubes [3], etc. However, reliable commercial products are still far to come, and one of the main reasons is the lack of long-term emission stability. The fundamental factors that contribute to the emission stability have not been well studied. Most reports have focused on low emission threshold fields of CNTs [4-9], electric field shielding effects [7], device architecture [8], and failure of CNTs after excessive current emission [9]. Practical cathodes with a high density of free-standing aligned CNTs, however, suffer from non-uniform current distributions and mutual shielding effects. Recently, degradation of current stability in multi-walled CNTs (MWCNTs) has been reported and explained also in terms of current-induced dislocation due the lower graphitic orders [10]. Moreover, pressure dependent current fluctuations and limits leading to insufficient long-term stability have been observed. In the literature, FE investigations are mostly carried out in a parallel plate setup, with diode or triode configuration, where current is averaged on a large sample area. But traditional I–V measurements with large area anodes are insufficient for proper FE characterization of densely packed CNT films. They are normally composted of few high $\beta$ emitters ($\beta \sim 1000$) and a majority of low $\beta$ emitters ($\beta \sim 100$). By exploiting a large area anode the emission is usually dominated by a comparatively small number of very strong emitting sites while emitters with a lower aspect-ratio (and also a lower field amplification factor $\beta$) are not detected. Moreover, since strong emitters represent a very small fraction of all possible emitters, the emission is not homogeneous and the obtained current densities are low. On the other hand, by probing the FE properties of small areas (less than 1μm$^2$) it is easy to find location where no strong emitters are present and the multitude of emitters with lower β value can be probed.

In this paper, we report the study of the long-term stability of the field emission current from vertically aligned MWCNTs performed by means of an atomic force microscope. The use of a metallic cantilever as anode gives access to small area properties and thus to the behaviour of emitters with lower β, that in the usual large anode configuration are not detected due to the presence of few (long) strong emitters usually protrusing with respect other tubes. We experimentally proved the importance of long electrical conditioning to stabilize the FE properties of MWCNTs, completely removing adsorbates from nanotubes.

## 2. Experimental

The CNTs were synthesized by catalytic chemical vapor deposition on a silicon p-type wafer (001), covered by a thin (~3nm) layer of SiO$_2$, acting as a diffusion layer and preventing the formation of NiSi$_x$. A dense array of vertical and partially aligned carbon nanotubes has been obtained, tubes supporting each other by van der Waals forces. Morphological characterization by scanning electron microscope (SEM) and transmission electron microscope (TEM) revealed multi-walled CNTs with average length of 15μm, inner diameters of 5-10 nm and outer diameter of 15-25 nm. SEM imaging (Figure 1) showed a lost of verticality in the upper part of the sample, but this doesn't represent a limitation for suitable field emission applications since the nanotubes are typically stretched and aligned along the cathode-anode direction at a sensibly smaller voltage with respect the turn-on value [11,12].

The field emission measurements were carried out in a room temperature Omicron ultra high vacuum STM/AFM system, where a Pt-Ti coated cantilever probe with elastic costant $k \approx 300KHz$ was used as anode. The Ti-Pt coating consists of a 10-nm Pt layer on a 20-nm Ti sublayer, which increases adhesion and electromigration firmness of Pt. The Ti-Pt coating is formed on both tip and reflective side of the cantilever. Resulting tip curvature radius with the coating is less than 40 nm. The high resolution piezo-tube system permits precise control of tip-sample relative position, with accuracy better than 0.1nm in any (x, y, z) direction and allowing a maximum tip-sample separation $d$ of 2µm. A high resolution source-measurement unit (SMU, Keitley 4200-SCS) was used to apply voltage up to ±210 V between tip and sample and to measure current with accuracy better then 1pA. Our setup, exploiting a very pointed anode, enables FE measurements over a limited circular region, whose radius has been estimated to be about 500nm for a tip-separation distance d=1µm. Such estimation has been obtained by a numerical simulation of the electrical field considering both geometrical configuration and materials characteristics [13].

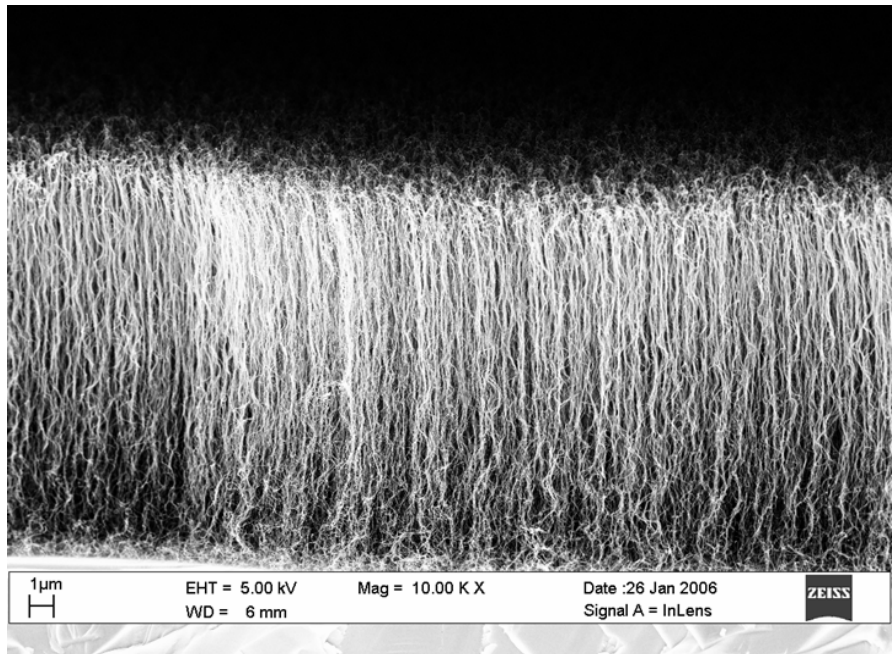

**Figura 1: (a) Lateral SEM image of the MWCNT film used for the field emission experiment.**

## 3. Results and discussions

In Figure 2 we report an example of current voltage characteristic measured in high vacuum (P=2.0·10$^{-8}$ mbar) by applying a bias voltage up to 210V to the AFM tip. We systematically observed that the initial electrical sweeps on virgin zones have a positive conditioning effect: irreversible changes, resulting in a stabilization, were found on the I-V characteristics. During the first sweep, the emission started between 30 V and 40 V, but for increasing voltage we observe that the current suffers abrupt drops, up to one order of magnitude. We precise that contact currents could be distinguished from the FE current by an abrupt current raise and saturation of the source-measure device, that we normally set at 10$^{-4}$ A to prevent possible burning of CNTs. Indeed, for such high currents, we often observed dramatic failure of the emission. This irreversible degradation probably happens due to a resistive heating of the MWCNTs [14].

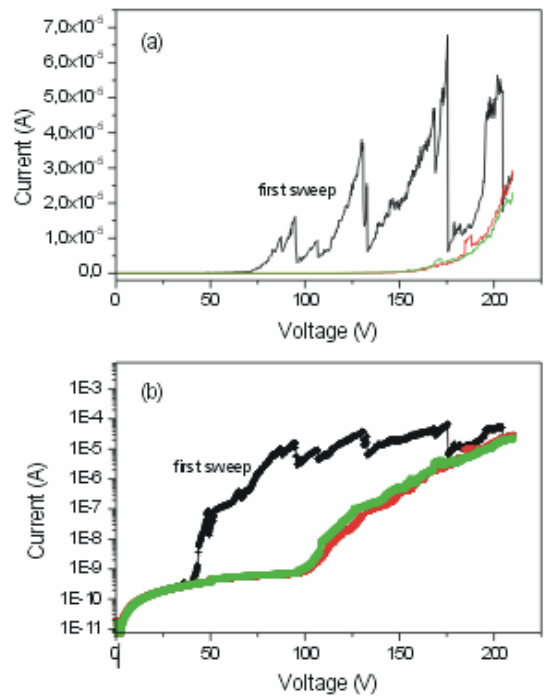

**Figura 2: Field emission current versus bias voltage in (a) linear and (b) logarithmic scale measured at 1μm chatode-anode separation and pressure of 2·10$^{-8}$ mbar. After a first sweep from 0 to 200 V the I-V characteristic remains stable for successive sweeps.**

Normally, after the first sweep FE becomes more stable; rarely the stabilization process requires additional sweeps. Following sweeps show a different behaviour, with higher turn-on voltage, about 100 V, and considerably lower current (the higher the voltage the lower the current difference).

The electrical conditioning observed for first sweep can be ascribed either to presence of few longer nanotubes with dominant FE that are gradually degraded till complete destruction for increasing current, or to desorption of adsorbates caused by CNT heating, topography changes due to CNT stretching and re-orientation, particulate cleaning, etc. In particular, adsorbates [15,16], as different types of gases, are always present at the CNT surface, creating nanoprotrusions i.e regions of reduced workfunction and increased enhancement factor, where field emission begins at lower electric fields. The formation and the electric field-driven surface diffusion of those nanoprotrusions can cause the observed instabilities of the FE current. At large current values, the local temperature becomes high enough to evaporate some of the adsorbates, provoking drops in the FE current.

Field emission is usually analyzed using Fowler–Nordheim (FN) theory, according to which the emission current I is mainly determined by the local electric field ($E_S$) at the emitter surface and by the work function ($\varphi$) of the material, as expressed by the equation $I \propto \left(E_S^2/\varphi\right)\exp\left[b\varphi^{3/2}/E_S\right]$ where b=6.83·10$^7$ eV$^{-3/2}$Vcm$^{-1}$ is a constant. By plotting experimental values of electron emission current versus applied voltage in the so-called Fowler–Nordheim plot, i.e., ln($I/V^2$) versus $1/V$, one should obtain a straight line with slope $m = \left(b\varphi^{3/2}d\right)/\beta$ (this formula is valid for standard parallel plate configuration) where β is the field enhancement factor that takes into account the amplification occurring around the apexes of the CNTs in the film. The FN plot reported in Figure 3 refers to one of the successive sweeps (after the conditioning) in Figure 2. The linear behavior confirms that the measured currents are field emission currents. From the numerical fitting the slope in the FN plot results to be $m = (1772 \pm 7)$ V. In evaluating the field enhancement factor β from the slop *m* we need to consider the geometry of our setup in which the anode is also tip shaped. This makes necessary to introduce a further enhancement factor that depends from the tip shape through the curvature radius of the apex and a so-called tip correction factor *k*. It has been shown [13] that the formula of the slope in the FN plot can be reasonably approximated by $m = (k_{eff} b \varphi^{3/2})/\beta$ where $k_{eff} \approx 1.6$ for our geometry, and consequently we can easily evaluate $\beta \approx 70$.

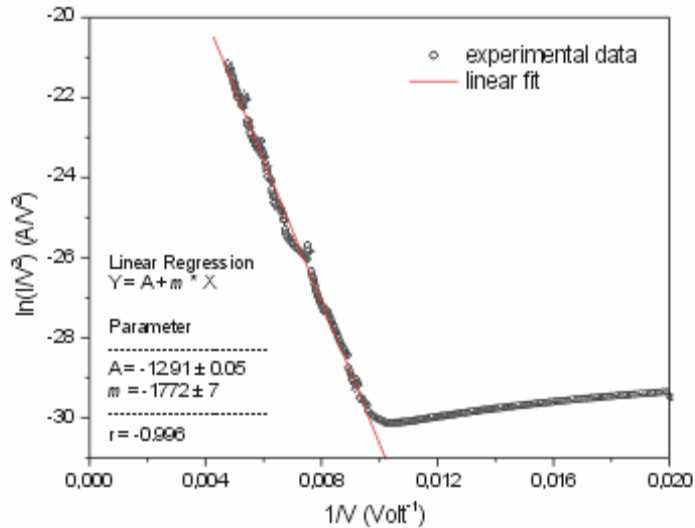

**Figura 3: F-N plot for one of the successive sweeps after electrical conditioning reported in Figure 2. Parameter for the linear fit are listed. A correlation factor r=0.996 indicates a very good fit.**

The use of a metallic local probe (AFM tip) as electrode for collecting electrons from the CNT film in the field emission experiments gives access to the study of the FE properties (I-V and stability) on a scale (≤1 µm$^2$) well below what is usually analyzed by the standard parallel plate configuration (useful to probe properties on areas from several tens of µm$^2$ to

mm$^2$). By scanning the probe over the surface, we had the possibility to measure the current-voltage characteristic in hundreds different locations either to have a statistical analysis of the homogeneity of sample behaviour and to verify the presence of small areas with degraded field emission properties.

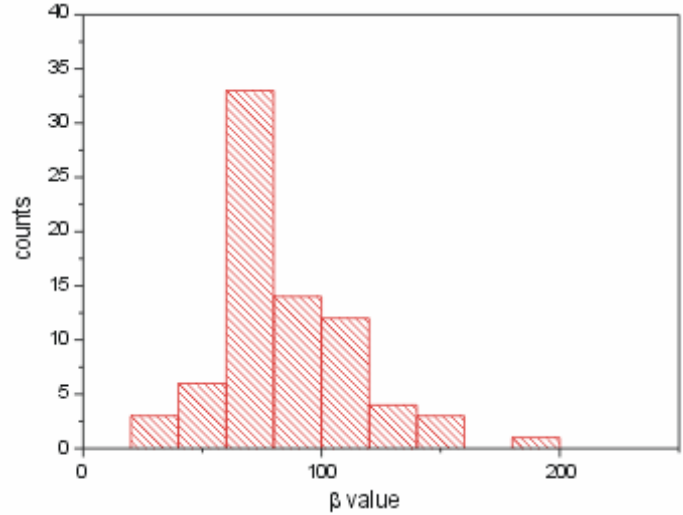

**Figura 4: Istogram of the β values recorded in several different locations of the MWCNT film.**

The result of a such large amount of data is summarized in figure 4, where we report an histogram of the β values obtained in several different locations. From statistical analysis it results that more than 40% of location show $60 < \beta < 80$, while 80% of locations show $60 < \beta < 120$. These data confirm the suitability of a nanometric probe for analyzing FE properties on enough small areas avoiding contributions from the rare strong emitters, that however are present as confirmed by the few cases in which we measured much higher β values. Also the values measured for the field enhancement factor are in agreement with the consideration that significant field screening effect (the applied electric field on the apex of a CNT is screened from the neighbouring CNTs [17]) is expected on our sample because of the high density of CNTs (it has been reported that the optimal condition for high emission is a tube spacing twice the CNT length [7]).

The field enhancement factor is known to be an increasing function of the inter-electrode distance and, taking into account the short distance $d$ (0.5-2 μm) used in our experiment, we get a rather high values of β, likely due to the conspicuous length of our nanotubes. Published values range from several hundreds to few thousands, with inter-electrode distances orders of magnitude bigger than in our case [18].

*3.1 Time Stability of the local field emission current*

We noted that carbon nanotubes show excellent field emission performances. Although the operation voltage is an important parameter for applications, the stability of emission current density is essential. Degradation of the emitting performances can be evaluated by measuring the evolution with time of emission intensity at constant applied voltage. We have carried out the comparative studies on many locations of our MWCNT films for different current intensities.

In figure 5 we report an example of the field emission properties, i.e. the current-voltage curve (I-V) and the current stability, in a location of the vertically aligned MWCNT film. Figure 5a shows a comparison between the I-V measured before (continous black line) and after (scattered red dots) the stability test (figure 5b) performed at fixed voltage $V_1$= 140V ($I = 10^{-6}$ A). We notice that the field emission current is quite noisy with oscillations of about 10% over few hours and that the first I-V curve measured after the stability test evidences a lower turn on voltage indicating a still not really complete conditioning of the area under the tip.

Again Figure 5c shows a comparison between the I-V dependences measured before (continous black line) and after (scattered red dots) the stability test (figure 5d) performed at a higher fixed voltage $V_2$= 160V ($I = 10^{-5}$ A), evidencing again a modification of the I-V curve after long term stability although the field emission current results less noisy but still showing a rising behavior in the initial test period of about 1 hour. In addition to this, we observe that the long test (about 50 h) had further important conditioning effects on the MWCNT film area under the tip. Indeed, from figure 5e and 5f we clearly see that I-V characteristics measured before and after the stability test (figure 5f) are identical, suggesting that the CNTs have reached stable configuration, not modified also by long emission periods and high current values.

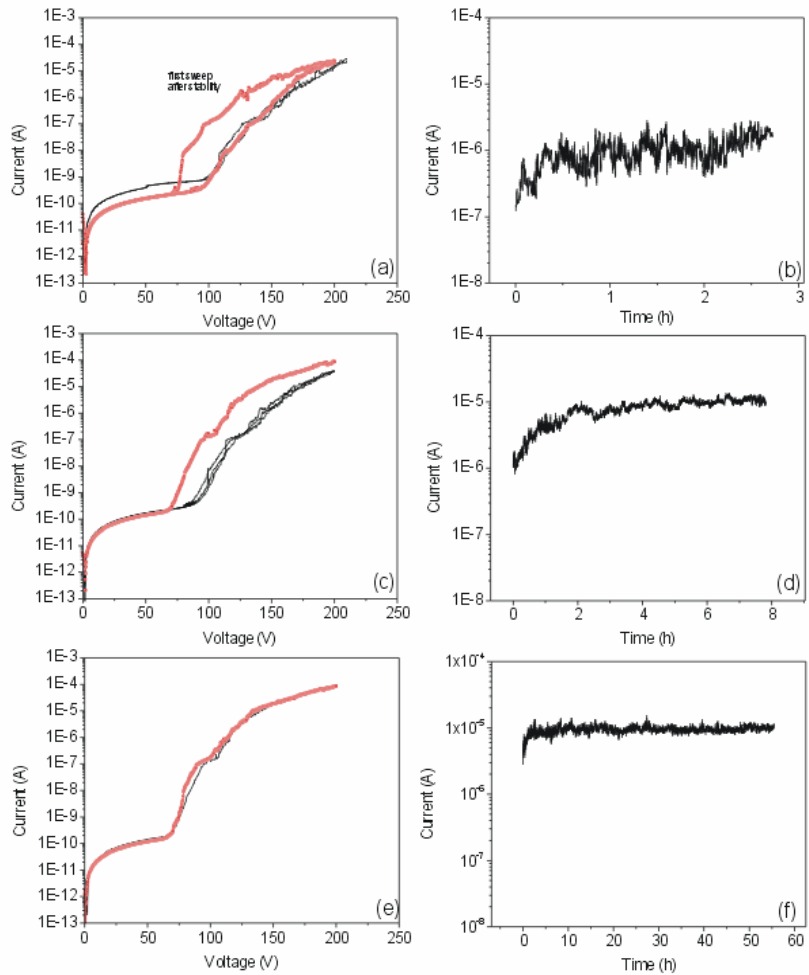

**Figura 5: Effect of long Joule heating on I-V characteristic and current stability. (a) I-V curves measured before (black) and after (red) first stability test at I=10$^{-6}$A (b); (c) I-V curves measured before (black) and after (red) stability test at I=10$^{-5}$A (d); (e) I-V curves measured before (black) and after (red) stability test at I=10$^{-5}$A (f).**

We already discussed that conditioning effect during current-voltage sweeps can be due to the presence of adsorbates, causing noise and flickering in the emission current and a relatively low threshold fields as usually observed during the first cycle. Then, stable I–V curves are reproducibly obtained, that again can be explained in terms of evaporation of the adsorbates by Joule heating during field emission [19,20]. However, we cannot completely ignore the possibility of residual adsorbates on the MWCNTs. From our data, we infer that only a long stability test at high current can realize a joule heating able to completely remove the adsorbates and to stabilizy definitively the I-V behavior and the emission current, as observed in figures 5e and 5f, respectively.

A confirmation of the improvement of the current stability obtained after long conditioning procedure consists in repeating the stability tests for different current values, as reported in figure 6 where we show measurements of emission current over a period of about 20 hours at the current values of (a) I= $1.0 \cdot 10^{-7}$A; (b) I= $2.5 \cdot 10^{-6}$ A; (c)=$1.0 \cdot 10^{-5}$ A. In all cases oscillations of the current remains below 10%.

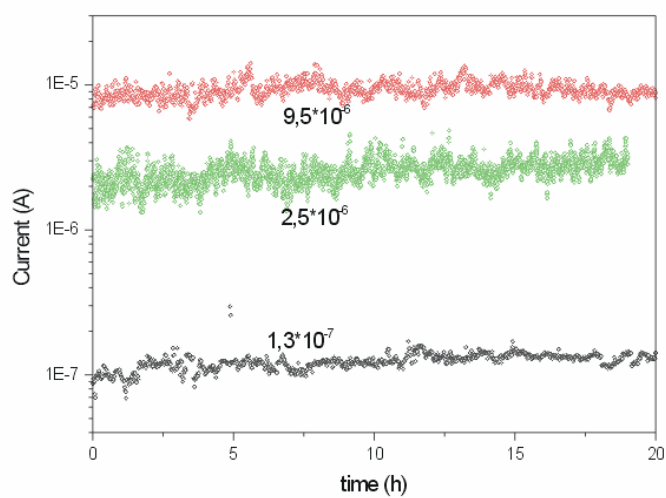

**Figura 6: Field emission current stability measured for different intensities of the emitted current on the same area.**

A further example of the performance of a long conditioning process at high emission current (I=$10^{-5}$A) is reported in figure 7 for one of the several virgin regions of the MWCNT film on which the conditioning procedure has been tested. We compare the current stability at $10^{-7}$ A measured before (figure 7b) and after (figure 7d) performing a 20 hours conditioning process at $10^{-5}$ A (figure 7c). In figure 7a we show the FE current-voltage characteristics: first sweep (black scattered dots) causes the usual conditioning with low (about 30V) turn on field, high current values and several drops in the current; an other sweep performed before any stability measurement shows stable behavior with higher turn on field (about 90 V); the red continous line represents the first sweep recorded after a stability test at $I=10^{-7}$ A, still evidencing a modification in the I-V curve. Only after the long stability at $I=10^{-5}$A (figure 7c) we get a stable I-V characteristic (blu line+dot in figure 7a) and high current stability for more then 70 hours with low noise level (<10%).

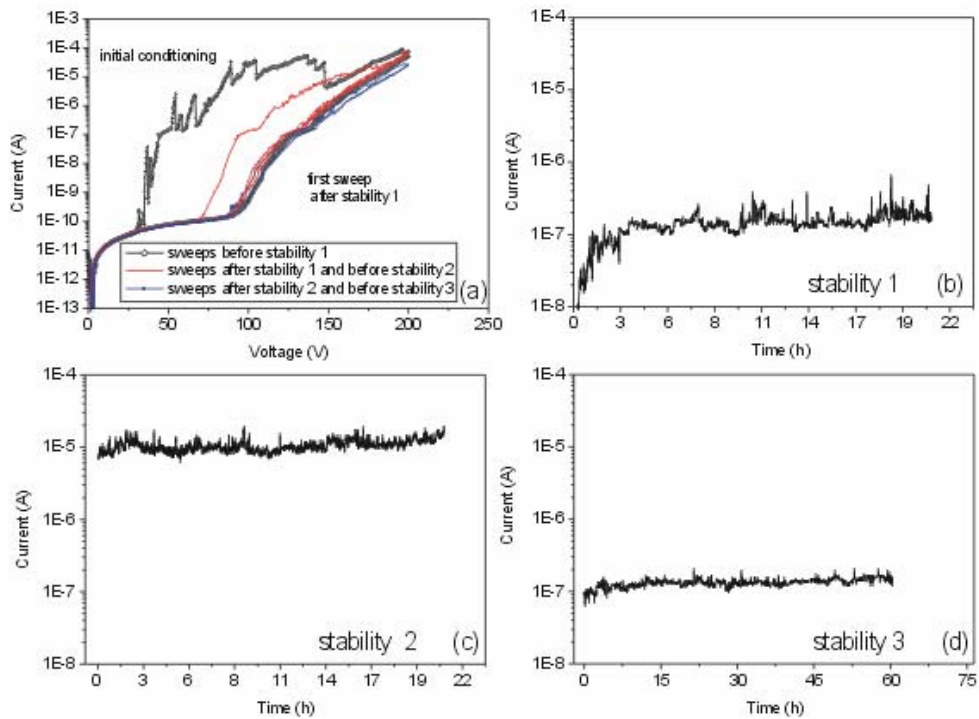

**Figura 7: Effect of Joule heating on the field emission current stability in one virgin area. (a) I-V curves; (b) first stability test for emitted current with intensity of $10^{-7}$A; (c) long time Joule heating by means of emitted current at $10^{-5}$A; (d) repetition of the stability test for emitted current with intensity of 10-7A after the long Joule heating.**

We notice that the usual conditioning effect due to a first current-voltage sweep in a virgin location is normally effective on adsobartes but our results indicate that the short time (from few milliseconds to few seconds) during which the joule heating acts on the CNTs its not enough to have a definitive effect, probably not removing completely the adsorbates. Only a long time joule heating at high current completely removes the adsorbates realizing stable conditions resulting in stable and reproducible voltage sweeps and in very good performances in emission stability with deviation well below 10%.

In conclusion, we have performed local field emission experiments on MWCNT films by means of metallic coated AFM probes. The small electrode used to collect electrons from CNTs allowed to analyze areas as small as $1\mu m^2$ avoiding the contribution from the rare long strong emitters that often affect measurements on larger areas by standard parallel plate setup. Statistical analysis evidenced a distribution of the enhancment factor $\beta$ centered around $\beta=70$. Moreover, a detailed study of the current stability at different current levels showed the importance of long term Joule heating to obtain areas characterized by very stable emission properties.

**Acknowledgements**

We thank GINT (Gruppo INFN per le NanoTecnologie) collaboration for motivating and supporting this work.